\documentclass[sigconf]{acmart}

\usepackage{booktabs} 
\usepackage{CJK}
\usepackage{subfigure}
\usepackage{sistyle}
\SIthousandsep{,}
\usepackage{makecell}
\usepackage{multirow}
\usepackage{enumitem}
\usepackage{filecontents}
\usepackage{bbm}
\usepackage{algorithm}
\usepackage{algorithmic}
\usepackage{pgfplots}
\usetikzlibrary{patterns}
\AtBeginDocument{%
  \providecommand\BibTeX{{%
    \normalfont B\kern-0.5em{\scshape i\kern-0.25em b}\kern-0.8em\TeX}}}

\copyrightyear{2020}
\acmYear{2020}
\setcopyright{acmcopyright}\acmConference[CIKM '20]{Proceedings of the 29th ACM International Conference on Information and Knowledge Management}{October 19--23, 2020}{Virtual Event, Ireland}
\acmBooktitle{Proceedings of the 29th ACM International Conference on Information and Knowledge Management (CIKM '20), October 19--23, 2020, Virtual Event, Ireland}
\acmPrice{15.00}
\acmDOI{10.1145/3340531.3412047}
\acmISBN{978-1-4503-6859-9/20/10}

\makeatletter
\g@addto@macro\normalsize{%
  \abovedisplayskip 2pt plus1pt 
  \belowdisplayskip 2pt plus1pt
  \abovedisplayshortskip  2pt plus1pt%
  \belowdisplayshortskip  2pt plus1pt
}
\makeatother

\begin{document}

\title{Continual Domain Adaptation for Machine Reading Comprehension}

\author{Lixin Su, Jiafeng Guo, Ruqing Zhang, Yixing Fan, Yanyan Lan, and Xueqi Cheng}
\affiliation{
  \institution{
    CAS Key Lab of Network Data Science and Technology, Institute of Computing Technology, \\ Chinese Academy of Sciences, Beijing, China\\
    University of Chinese Academy of Sciences, Beijing, China\\} 
}
\email{sulixinict@gmail.com, {guojiafeng,zhangruqing,fanyixing,lanyanyan,cxq}@ict.ac.cn}

\renewcommand{\shortauthors}{Su and Guo, et al.}

\begin{abstract}

Machine reading comprehension (MRC) has become a core component in a variety of natural language processing (NLP) applications such as question answering and dialogue systems. It becomes a practical challenge that an MRC model needs to learn in non-stationary environments, in which the underlying data distribution changes over time. A typical scenario is the domain drift, i.e. different domains of data come one after another, where the MRC model is required to adapt to the new domain while maintaining previously learned ability. To tackle such a challenge, in this work, we introduce the \textit{Continual Domain Adaptation} (CDA) task for MRC. So far as we know, this is the first study on the continual learning perspective of MRC. We build two benchmark datasets for the CDA task, by re-organizing existing MRC collections into different domains with respect to context type and question type, respectively. We then analyze and observe the catastrophic forgetting (CF) phenomenon of MRC under the CDA setting. To tackle the CDA task, we propose several BERT-based continual learning MRC models using either regularization-based methodology or dynamic-architecture paradigm. We analyze the performance of different continual learning MRC models under the CDA task and show that the proposed dynamic-architecture based model achieves the best performance.

\end{abstract}

%

%

\maketitle

\section{Introduction}
Machine Reading Comprehension (MRC) aims to read and understand unstructured texts and then answer questions about it, which has become a core component in a variety of natural language processing (NLP) applications \cite{chen2018neural}. For example, in retrieval-based question answering systems, e.g., Web QA \cite{chen2017reading}, MRC plays a vital role which attempts to comprehend the retrieved documents to extract a concise answer for a given question. In dialogue systems, e.g., conversational information assistant \cite{Qin2019ConversingBR}, MRC is often employed as a basic component to build professional response skills over some target information resources such as medical or legislative corpus.

 \begin{figure}[t]
 \centering
 \includegraphics[scale=0.36]{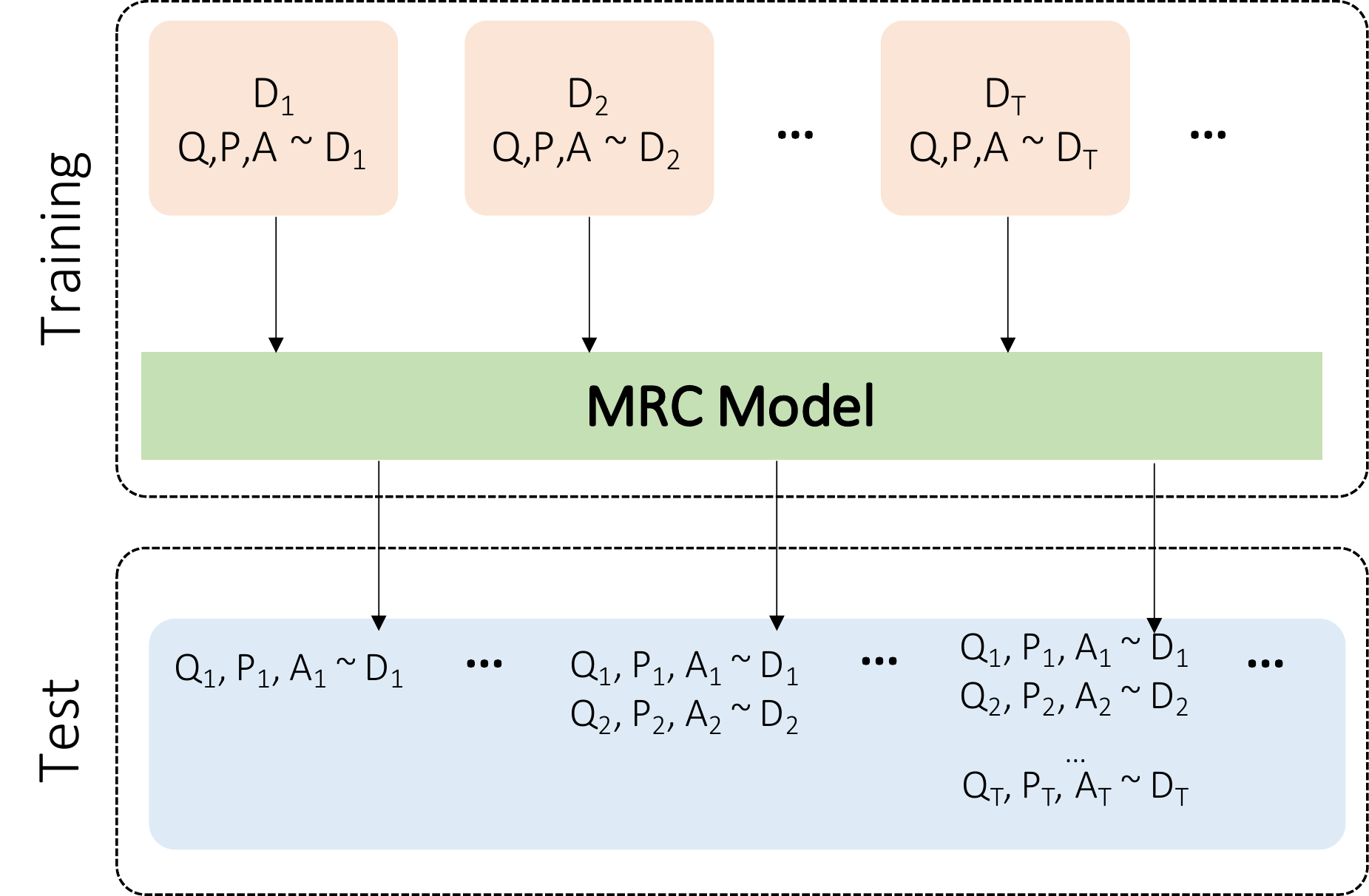}
 \caption{Illustration of continual domain adaptation task for machine reading comprehension.}
 \label{fig:intro}
 \end{figure}
 
As a component of real-world applications, it becomes a practical challenge that an MRC model needs to learn in non-stationary environments. That is the underlying data distribution where a model is applied changes over time. A typical case is called \textit{domain drift} \cite{Lao2020ContinuousDA}, where different domains of data come one after another. For example, a question answering system may gradually enlarge its retrieved repository from news articles to social media, or from factoid information to non-factoid information. A dialog system, e.g. a medical information assistant, may consistently expand its skill to provide information about more and more types of diseases. These practical issues require the underlying MRC model to adapt to the new domain while maintaining previously learned knowledge. Note that in practical scenarios, an early domain of data may not be available in a later phase due to some privacy or storage issues.

To study such practical challenge, in this work, we introduce the \textit{Continual Domain Adaptation} (CDA) task for MRC. In this CDA task as shown in Figure~\ref{fig:intro}, training data from different domains arrives in a sequential manner, and an MRC model is required to continually learn over each domain. Evaluation of CDA is to test the overall performance on all the domains that the MRC model has encountered. As we knew, there have been a large variety of MRC tasks \cite{fisch2019mrqa} proposed in the literature. However, all those tasks assume a stationary learning scenario, i.e., a fixed data distribution. So far as we know, this is the first task about the continual learning ability of MRC models.

To facilitate the study of the CDA problem, we build two benchmark CDA datasets for MRC, namely CDA-C and CDA-Q, by re-organizing existing MRC collections into different domains with respect to context type and question type, respectively. For the CDA-C dataset, we make use of five MRC collections including  SQuAD \cite{Rajpurkar2016SQuAD10}, NewsQA \cite{Trischler2017NewsQAAM}, TweetQA \cite{Xiong2019TWEETQAAS}, NarrativeQA\cite{Kocisk2018TheNR}, and DuoRC \cite{Saha2018DuoRCTC}. We obtain a 5-domain dataset to mimic the drift of the context distribution, from wiki passages, news articles, movie scripts, book summaries to tweets. For the CDA-Q dataset, we make use of the NaturalQA \cite{Kwiatkowski2019NaturalQA} collection and split it into 8 domains according to the question types such as what, why, and how. Both datasets would be publicly available for the research community\footnote{https://github.com/lixinsu/CDA-CIKM2020}. Based on these two CDA datasets, we conduct some empirical analysis over the existing MRC models to check whether the MRC model suffers from the domain drift problem. We employ a state-of-the-art MRC model, i.e., BERTQA \cite{Devlin2019BERTPO}, for the analysis. We observe the very obvious forgetting phenomenon of the BERTQA model under the CDA setting~\ref{sec:cf} similar to \cite{Li2018LearningWF}, i.e., a general tendency to forget past knowledge as it learns over new domains.

To tackle the CDA task for MRC, we propose to equip the existing state-of-the-art MRC models with the continual learning ability. Continual learning is an exciting direction in the machine learning community, which has been actively studied in the computer vision (CV) area in recent years \cite{Ostapenko2019LearningTR,Li2018LearningWF,Schwarz2018ProgressC,Lee2017OvercomingCF} but received less attention in NLP. The exiting continual learning techniques could be categorized into two major branches, i.e., regularization-based methods and dynamic-architecture methods. Regularization-based methods \cite{Kirkpatrick2017OvercomingCF} identify important weights for previous domains and heavily penalize their deviation while learning new domains with some type of regularization. 
Dynamic-architecture methods \cite{rusu2016progressive} append new components or eliminate learned weights in the network when learning with continually arrived domains.

We borrow the idea from the CV area and propose several BERT-based continual learning MRC models using either regularization-based methodology or dynamic-architecture paradigm. 
Based on the regularization-based methodology, we propose RegBERTQA which adds a penalty to the learning objective of the BERTQA model which restricts the change of parameters to prevent forgetting. Based on the dynamic-architecture paradigm, we introduce a novel ProgBERTQA model for MRC, which enlarges its model capacity progressively when a new domain arrives. Specifically, the ProgBERTQA model leverages the BERT model as a backbone structure and adds an adapter component for each incoming domain. The adapter is learned for the current domain while the backbone parameters which are domain-shared are kept unchanged all the time. We studied two types of adapter architectures and two ways to insert the adapter into the BERT model. We conduct extensive experiments to analyze the performance of different continual learning MRC models under the CDA task. Based on the empirical results, we find that the ProgBERTQA, which adopts the dynamic-architecture paradigm, achieves the best performance on the CDA task for MRC.

The major contributions of this paper include: 

\textbf{1.} We introduce the CDA task, which is the first task focusing on the continual learning ability of MRC models, and show that existing MRC models do suffer from the CF phenomenon under the CDA setting.

\textbf{2.} We build two publicly available benchmark datasets that could facilitate the study of the CDA problem of MRC in the future.

\textbf{3.}  We propose and investigate several BERT-based continual learning MRC models, which could combat the CF phenomenon while learning over new domains continually.

\section{Related Work}
In this section, we first briefly review two lines of related work, i.e., machine reading comprehension and continual learning. 
Then, we introduce various datasets for machine reading comprehension research in detail.

\subsection{Machine Reading Comprehension}
Teaching machines to read and comprehend is a fundamentally interesting and challenging problem. MRC plays a vital role in a question answering system for comprehending retrieved passages to achieve a concise answer~\cite{chen2017reading}. 
Existing literature on MRC models has focused on the standard supervised setting,  where the training data and the test data are assumed to sample from the same distribution.
BiDAF \cite{Seo2017BidirectionalAF} takes RNN for representing the context and the question. Consequently, bidirectional attention flow mechanism is applied to obtain a query-aware context representation. This context representation is further used to predict the start position and the end position of the answer.  
FusionNet \cite{Huang2018FusionNetFV}  also applies RNN for text representation and proposes a fully-aware multi-level attention mechanism to capture hierarchical information of each word.  
QANet \cite{Yu2018QANetCL} applies the structure consisting of convolution and self-attention for fast training. Convolution layer captures local interactions, and self-attention mechanism perceives global interactions.
BERT \cite{Devlin2019BERTPO} is a self-supervised approach for pre-training a deep transformer encoder \cite{Vaswani2017AttentionIA}. BERTQA is a fine-tuned BERT for the MRC task and achieves state-of-the-art performance. 
In this paper, different from existing MRC models designed for the stationary dataset, our goal is to propose an MRC model that can handle domain drift to achieve continual domain adaptation.

\subsection{Continual Learning}
Continual learning aims to sequentially learn a series of tasks.
Existing methods for continual learning can be divided into two main categories: network regularization and dynamic architecture. 
In the following, we give a brief review of typical methods in two categories.

\subsubsection{Network regularization} 
Regularization-based approaches suggest using a regularizer that prevents the parameters from drastic changes in their values yet still enables them to converge to a good solution for the new task \cite{Chaudhry2018RiemannianWF, Dhar2019LearningWM,Kirkpatrick2017OvercomingCF,Lee2017OvercomingCF,Lee2017OvercomingCF}. 
To keep the learned knowledge, various penalties are added to the change of weights. EWC \cite{kirkpatrick2017overcoming} uses Fisher’s information to evaluate the importance of weights for old tasks, and updates weights according to the degree of importance. Based on similar ideas, the method in \cite{Zenke2017ContinualLT} calculates the importance based on the learning trajectory. Online EWC \cite{Schwarz2018ProgressC} and EWC++ \cite{Chaudhry2018RiemannianWF} improve the efficiency issues of EWC. The approach builds an attention map and requires that the attention region of the previous and concurrent models are consistent. 
There are related methods \cite{Shin2017ContinualLW} using additional models to remember previous data distribution to regularize the learning process. 
Generative Replay \cite{Shin2017ContinualLW} introduces GANs to lifelong learning, which uses a generator to sample fake data. New tasks can be trained with these generated data. 

These regularization-based methods can be applied to different model structure, and we adapt representative methods EWC to MRC model for CDA.  

\subsubsection{Dynamic Architecture} Dynamic-architecture approaches \cite{Li2018LearningWF} dynamically increase model capacity during training. ProgressiveNet \cite{rusu2016progressive} expands the architecture for new tasks and keeps previous knowledge by preserving the previous weights. LwF \cite{Li2018LearningWF} divides the model layers into two parts, i.e., shared and task-specific layers. The former are shared by tasks, and the later are expanded for new tasks. DAN \cite{Collazo2018TowardAR} extends the architecture for each new task, and the newly added layer is a sparse linear combination of the original filters in the corresponding layer of a base model. 
Asghar et al.~\cite{asghar2018progressive} augment parameterized memory for sequentially arrived tasks in RNN structure for text classification task.

These dynamic methods cannot be applied to MRC model, as they are design for CV models or RNN structure. Inspired by these dynamic architectures, we aims to design a dynamic MRC model based pre-trained transformer for CDA task.

\subsection{MRC Datasets}

There are many released MRC datasets for developing and evaluating MRC models, which can be generally categorized into basic datasets and derived datasets.

\subsubsection{Basic Datasets}
MRC datasets can be categorized into three types with respect to questions. Cloze-style MRC datasets~\cite{Hermann2015TeachingMT} aims to teach machines to read the passage and fill the blank in questions. Multiple-choice MRC datasets~\cite{Lai2017RACELR} give the context, the question, and several candidate answers, then asks models to choose the correct answer.  Extractive MRC \cite{Rajpurkar2016SQuAD10} aims to locate text span from the context given the context and the question.
Extractive MRC has attracted extensive attentions, since it can be used in real question answering system and question answering system \cite{Qin2019ConversingBR}.
In this paper, we focus on extractive MRC and we ignore ``extractive'' for brevity. We then describe some details of MRC datasets.
MRC datasets such as SQuAD, NarrativeQA, and HotpotQA contain question-passage-answer triple.  Questions are written by annotators after they read context passages, and answers are short text spans in the passages. 
SQuAD~\cite{Rajpurkar2016SQuAD10} are constructed based on Wikipedia passages. 
NewsQA \cite{Trischler2017NewsQAAM} are constructed based on CNN news.
NarrativeQA \cite{Kocisk2018TheNR} use the passages from fictional story books, which exposes additional challenge for longer passages. 
NaturalQA \cite{Kwiatkowski2019NaturalQA} is a more natural dataset, as its questions consist of real anonymized, aggregated queries issued to the Google search engine and its context passages are Wikipedia page from the top 5 search results. NaturalQA effectively eliminates the plagiarism from passage to question as question is collected before contexts. 
TweetQA \cite{Xiong2019TWEETQAAS} presents the first large-scale dataset for QA over social media data.
It gathers tweets as context passaegs and human annotators write questions and answers upon these tweets.
DuoRC \cite{Saha2018DuoRCTC} is constructed from a collection of movie plot pairs where each pair in the collection reflects two versions of the same movie written by different authors. questions are generated based on one version and answers are synthesized from the other version.  

\subsubsection{Derived Datasets}
Given extensive MRC datasets, some works reuseed those datasets to comprehensively test the high-level ability of models. 
GLUE~\cite{wang2018glue} collected a series of NLU tasks including question answering, sentiment analysis, and textual entailment. 
GLUE is a multitask benchmark in NLP, designed to encourage models that share general linguistic knowledge across tasks.
Wang et al.~\cite{wang2019adversarial} grouped SQuAD, NewsQA, and MS MARCO to study the unsupervised domain adaptation for MRC. They useed a full  MRC dataset as the source domain and take only unlabeled passages in another dataset as the target domain.
With this derived dataset, they test the model capability of unsupervised domain adaptation.
Fisch et al.~\cite{fisch2019mrqa} presented Machine Reading for Question Answering (MRQA) 2019 shared task, which tested extractive MRC  models on their ability to generalize to data distributions different from the training distribution. 
They unified 18 distinct question answering datasets into the uniform format. Among them, six datasets were made available for training, six datasets were made available for development, and the final six were hidden for final evaluation.

Different from existing derived datasets, we aim to construct a continual domain adaptation setting. We split existing datasets into different domains, and each domain is available to the MRC model sequentially.

\section{Continual Domain Adaptation for MRC}

In this section, we introduce the Continual Domain Adaptation (CDA) task for MRC and describe the benchmark CDA datasets for MRC in detail.

\subsection{Task Formulation}

In the CDA task for MRC, we assume that there exists $T$ MRC datasets $\{ \mathcal{D}_1, \dots, \mathcal{D}_T \}$ from $T$ different domains arriving in a sequential manner. 
We denote the dataset $\mathcal{D}_t$ from domain $t$ as, 
\begin{equation}
	\mathcal{D}_t = \{c^i_{t},q^i_{t},a^i_{t}\}_{i=1}^{N_t}
\end{equation}
where $c^i_{t},q^i_{t}$ and $a^i_{t}$ denotes the $i$-$th$ context, question and answer among $N_t$ samples.  
In the CDA setting, an MRC model is required to continually learn over each domain dataset $\mathcal{D}_t$, and evaluated  on all seen domains. 
Each time, the MRC model can only access current domain and cannot access previous training domains due to the privacy or storage issues.

The training objective at the arrival of domain $t$ can be defined as follows, 
\begin{equation}
minimize_{\Theta_t} \mathcal{L}(\Theta_t;\Theta_{t-1}, \mathcal{D}_t) + \lambda \mathcal{R} (\Theta_t),	
\end{equation}
where $\Theta_t$ denotes the set of parameters for the MRC model at domain $t$, and $\mathcal{R}(\cdot)$ is a regularization term on the model parameters.

For the evaluation of CDA, we employ the overall performance on all seen domains that the MRC model has beed trained on, 
\begin{equation}
	\sum_{t} \sum_{q^i_{t},c^i_{t},a^i_{t} \in \mathcal{D}_t}  g(a^i_{t},f(q^i_{t},c^i_{t};\Theta_t)),
\end{equation}
where $f(\cdot)$ denotes the learned MRC model and $g(\cdot)$ denotes the widely used evaluation metric for MRC, e.g., EM and F1. 
The detail about evaluation will be described in section \ref{sec:eval}.

\begin{table}[]
\renewcommand{\arraystretch}{1.5}
   \setlength\tabcolsep{5pt}
\centering
\begin{tabular}{cccccc} \hline \hline
datasets   & wiki   & news   & scripts & book   & tweet  \\ \hline
\#train    & 10,000 & 10,000 & 10,000  & 10,000 & 10,000 \\
\#test      & 10,507 & 4,212  & 9,344   & 9227   & 4212   \\
\#q\_words & 10.2   & 6.5    & 7.4     & 8.4    & 6.5    \\
\#a\_words & 3.0    & 3.9    & 1.6     & 1.8    & 3.9    \\
\#c\_words & 124.0  & 492.6  & 658.2   & 580.1  & 492.6 \\ \hline \hline
\end{tabular}
\caption{Dataset statistic of CDA-C.}
\label{table:CDA-C}
\end{table}

\begin{table}[]
\renewcommand{\arraystretch}{1.5}
   \setlength\tabcolsep{1.5pt}
\centering
\begin{tabular}{ccccccccc}  \hline \hline
datasets   & what  & which & where & when  & how   & why   & other & who   \\\hline
\#train    & 16809 & 3052  & 12136 & 19470 & 5791  & 530   & 8114  & 38169 \\
\#test     & 2221  & 383   & 1501  & 2256  & 653   & 56    & 1014  & 4752  \\
\#q\_words & 9.3   & 10.6  & 8.7   & 8.9   & 9.3   & 8.8   & 9.8   & 9.1   \\
\#a\_words & 6.3   & 3.2   & 7.2   & 3.5   & 3.6   & 15.7  & 4.7   & 3.0   \\
\#c\_words & 153.7 & 168.5 & 126.4 & 158.6 & 166.6 & 113.8 & 156.4 & 151.7\\\hline \hline
\end{tabular}
\caption{Dataset statistic of CDA-Q.}
\label{table:query}
\end{table}

\subsection{Benchmark Construction}

In order to study and evaluate the CDA model for MRC, we build two benchmark datasets, i.e., CDA-C and CDA-Q. Datasets are constructed based on several existing MRC collections by splitting them into different domains with respect to context type and question type respectively.  
We now describe the detail of the CDA-C and CDA-Q dataset as follows.

\subsubsection{CDA Dataset for Context Drift}

To mimic the drift of the context distribution in the CDA task, we construct a new benchmark dataset called CDA-C. 
We take five MRC collections including SQuAD from Wiki passages \cite{Rajpurkar2016SQuAD10}, NewsQA from CNN news articles \cite{Trischler2017NewsQAAM}, TweetQA from tweets \cite{Xiong2019TWEETQAAS}, NarrativeQA from book summaries \cite{Kocisk2018TheNR}, and DuoRC from movie scripts \cite{Saha2018DuoRCTC} as our whole MRC datasets, since 
(1) These collections are publicly available and the size of each dataset is big enough for training deep neural models; 
(2) The contexts in these collections are different from each other and it is reasonable to distinguish one domain from another domain. 
Thus, we can obtain a 5-domain datasets to mimic the drift of the context distribution. 
Specifically, for saving computation cost, we randomly sample $10,000$ $<$question, context, answer$>$ triples from each MRC datsets as the training set, and the original test set is used for testing. 
Table \ref{table:CDA-C} shows the overall statistics of our CDA-C  benchmark dataset.

\subsubsection{CDA Dataset for Question Drift}

In order to mimic the drift of the question distribution in CDA task, we build a new benchmark dataset CDA-Q. 
Here, we take the NaturalQA~\cite{Kwiatkowski2019NaturalQA} dataset as our source data. 
The reason is that NaturalQA contains various types of questions, and the amount of questions are much larger than other datasets~\cite{Rajpurkar2016SQuAD10}. 
Specifically, we first select $7$ question words (such as ``what'', ``how'' and ``why'') to represent the question types. 
Then, to identify the question type for each question, we locate the  question word in the question if it can be found in the first three tokens, or the last word otherwise. 
Note a question would be defined as ``other'' type if there exists no question words in it. 
In this way, we obtain a 8-domain dataset to mimic the drift of the question distribution. 
Table \ref{table:query} shows the overall statistics of our CDA-Q benchmark dataset.

\begin{table}[t]
\centering
\renewcommand{\arraystretch}{1.5}
   \setlength\tabcolsep{0.5pt}
\begin{tabular}{llllll}\hline \hline
                 & {wiki} & news & scripts & book & tweet \\ \hline
wiki    & 80.01      & -       & -          & -      & -        \\
news    & 75.63(-5.5\%)      & 58.5       & -          & -       & -        \\
scripts & 71.08(-11.2\%)     & 44.35(-24.2\%)       & 67.33         & -        & -        \\
book    & 69.97(-12.5\%)     & 44.25(-24.4\%)       & 67.73(+0.6\%)          & 61.21       & -        \\ 
tweet   & 67.9(-15.1\%)      & 44.69(-23.6\%)      & 67.8076(+0.7\%)          & 58.51(-4.4\%)     & 85.13       \\ \hline \hline

\end{tabular}
\caption{F1 of BERTQA on CDA-C. Domain sequence is \emph{wiki>news>scripts>book>tweet}.}
\label{tab:ptype}
\end{table}

\section{Analysis of BERTQA model on CDA}\label{sec:cf}

In this section, based on our constructed CDA-C and CDA-Q benchmark dataset, we conduct some empirical analysis over the existing MRC model to investigate whether the MRC model suffers from the domain drift problem, i.e., a sequence of continuously evolving domains, which has been identified in many computer vision methods~\cite{kirkpatrick2017overcoming}. 

Great efforts have been made to develop MRC models for better effectiveness. A significant milestone is that the pre-trained BERTQA model \cite{Devlin2019BERTPO} has achieved state-of-the-art performance on several widely used MRC datasets~\cite{Reddy2019CoQAAC,Choi2018QuACQ}, and even exceeded the performance of human annotators on the SQuAD dataset~\cite{Rajpurkar2016SQuAD10}.
Hence, we employ this representative BERTQA model for the analysis, and the model detail is described in section \ref{sec:bert}. 

\begin{table*}[t]
\renewcommand{\arraystretch}{1.4}
   \setlength\tabcolsep{5pt}
\centering
\begin{tabular}{lllllllll} \hline \hline
      & what    & which   & where   & when    & how     & why     & other   & who     \\ \hline
what  & 67.46 & - & - & - & - & - & - & - \\
which & 61.5(-8.8\%) & 66.23 & - & - & - & - & - & - \\
where & 49.73(-26.3\%) & 57.35(-13.4\%) & 68.64 & - & - & - & -   & - \\
when  & 47.26(-29.9\%) & 50.48(-23.8\%) & 56.35(-17.9\%) & 78.59 & - & -  & - & - \\
how   & 51.54(-23.6\%) & 48.03(-27.5\%) & 53.89(-21.5\%) & 55.47(-29.4\%) & 73.07 & - & - & - \\
why   & 32.69(-51.5\%) & 26.98(-59.3\%) & 30.94(-54.9\%) & 32.98(-58.0\%) & 38.14(-47.8\%) & 57.08 & - & -  \\
other & 59.63(-11.6\%) & 65.79(-0.7\%)  & 61.66(-10.2\%) & 63.95(-18.6\%) & 66.04(-9.6\%)  & 34.76(-39.1\%) & 62.89 & -  \\
who   & 54.81(-18.8\%) & 57.95(-12.5\%) & 54.66(-20.4\%) & 65.29(-16.9\%) & 60.96(-16.6\%) & 46.23(-19.0\%) & 57.48(-8.6\%) & 81.40 \\ \hline \hline
\end{tabular}
\caption{F1 of BERTQA on CDA-Q. Domain sequence is \emph{what >which>where>when>how>why>other>who}.}
\label{tab:qtype}
\end{table*}

To evaluate the performance of BERTQA on the CDA datasets, we use pre-trained BERT~\footnote{https://github.com/google-research/bert}
trained on English Wikipedia, and  we finetune it when each domain arrives. 
The BERTQA parameters learned from previously seen domains will be continually fine-tuned to the new domain during training. 
We report F1 performance on each domain in the test CDA-C and CDA-Q dataset.

Table \ref{tab:ptype} shows the evaluation results on test CDA-C dataset.  
We can observe that BERTQA performance on all earlier domains significantly drops as new domains arrive. 
For example, the drop F1 in the wiki domain at the last step over that at the first step is about 12.2.
The main reason might be that BERTQA forgets how to solve old domains after being exposed to a new one due to interference caused by parameter updates.   
Table \ref{tab:qtype} shows the evaluation results on test CDA-Q  dataset. 
We can see consistency performance drop in CDA-Q as in CDA-C. 
We can also find that there exists performance fluctuation in CDA-Q due to the similarity between different question domains while the overall model performance on earlier question domains drops as new domains arrive. 

By analyzing how BERTQA performance on earlier domains change as new domain arrive, we can verify that the MRC model suffers from obvious catastrophic forgetting in domain drift setting. 
Therefore, it is necessary to efficiently mitigate catastrophic forgetting while continuously learn new domains under the CDA setting.

\section{Methods}
In this section, we introduce our proposed BERT-based continual learning MRC models. 
We first introduce the basic BERTQA model and our proposed RegBERTQA model based on the regularization-based methodology. 
We then describe  our proposed ProgBERTQA model based on the dynamic-architecture paradigm as well as the learning procedure. 

\subsection{BERTQA} \label{sec:bert}

The BERTQA model is formed by incorporating BERT~\cite{Devlin2019BERTPO} with one additional output layer for MRC. 
Specifically, BERT's model architecture is a multi-layer bidirectional Transformer encoder~\cite{Vaswani2017AttentionIA}  composed of a stack of identical layers, where each layer has a self-attention layer and a feed-forward network layer.
With the question and context representation vectors from BERT available, BERTQA applies an output softmax layer over all of the representations in the context to predict the starting position probabilities and the end position probabilities of the answer span. We now describe the self-attention and feed-forward network sub-layer in Transformer layer as follows. 

\subsubsection{Self-Attention} The Self-Attention sub-layer aims to capture global information through multi-head attention (MH) and a linear layer. 
Multi-head attention allows the model to jointly attend to information from different representation subspaces at different positions, where the attention weights are derived by the dot-product similarity between transformed representations. 
Concretely, the $i$-th attention mechanism ``head'' is:
$$Attention_{i}(\mathbf{h}_j) = \sum_{t}softmax \left( \frac{W_i^q\mathbf{h}_{j} \cdot W_i^k \mathbf{h}_t}{\sqrt{d/n}} \right) W_i^v\mathbf{h}_t$$
where $\mathbf{h}_j$ is a $d$ dimensional hidden vector for a particular sequence token. $W_i^q$, $W_i^k$ and $W_i^v$ are learned matrices of size $d/n \times d$, and thus each ``head'' projects down to a smaller subspace of size $d/n$. Finally the outputs of the $n$ attentions heads are concatenated together and passed to a linear transformation:
$$MH(\mathbf{h})=W^o[Attention_{1}(\mathbf{h}), \dots, Attention_{n}(\mathbf{h})]$$
 with $W^o$ is a $d \times d$ matrix. 
 The outputs are further passed to residual network and layer normalization. 
 We denotes the process as SA(.): 
 $$SA(\mathbf{h})=LN(\mathbf{h}+MH(\mathbf{h}))),$$
 where LN(.) is layer normalization \cite{Ba2016LayerN}, requiring $2d$ parameters.

\subsubsection{Feed Forward Network} The outputs of SA is passed to two-layer feed forward network (FFN). FFN is shown as
 $$FFN(\mathbf{h})=W_2 f(W_1 \mathbf{h}+b_1) + b_2,$$
where matrix $W_1$ has size $d_{ff} \times d$ and $W_2$ has size $d\times d_{ff}$, $f(\cdot)$ is a activation function,so overall we required $2dd_{ff}$ parameters from the FFN layer.
 
 Putting this together, a BERT layer, which we denote as BL(.), is a layer-norm applied to the output of FFN layer, with a residual connection.
 $$BL(\textbf{h})=LN(FFN(SA(\textbf{h}))+SA(\textbf{h}))$$
 We have $4d^2+2dd_{ff}$ total parameters from a BERT layer.  
 
\subsection{RegBERTQA}

Based on the regularization-based methodology in continual learning~\cite{Kirkpatrick2017OvercomingCF}, we propose the RegBERTQA model which adapts EWC \cite{Kirkpatrick2017OvercomingCF} method to BERTQA. 
The key idea is that it is reasonable to prevent forgetting by restricting the change of parameters to keep the network parameters close to the learned parameters of the old domain. 

Specifically, RegBERTQA updates model parameters based on the importance evaluated in previously seen domains. 
We denote parameters of BERT at  $t$ domain as  $\Theta_{t}$. When $t-th$ domain arrives, we fine-tune BERTQA and add a penalty, weighted distance between old parameters $\Theta_{t-1}$ and $\Theta_{t}$, denoted as $R(\Theta_{t}) = \sum_{i} F_i (\Theta_{t,i}-\Theta_{t-1,i})^2,$ where $F$ Fisher information matrix~\cite{Kirkpatrick2017OvercomingCF} which is a approximation of importance, i of $F_i$ denotes each parameter. Note that there are no penalty for the first domain. RegBERTQA need to maintain two set of BERTQA parameters, one for old parameters and one for active parameters.
	
 \begin{figure}[t]
 \centering
 \includegraphics[scale=0.4]{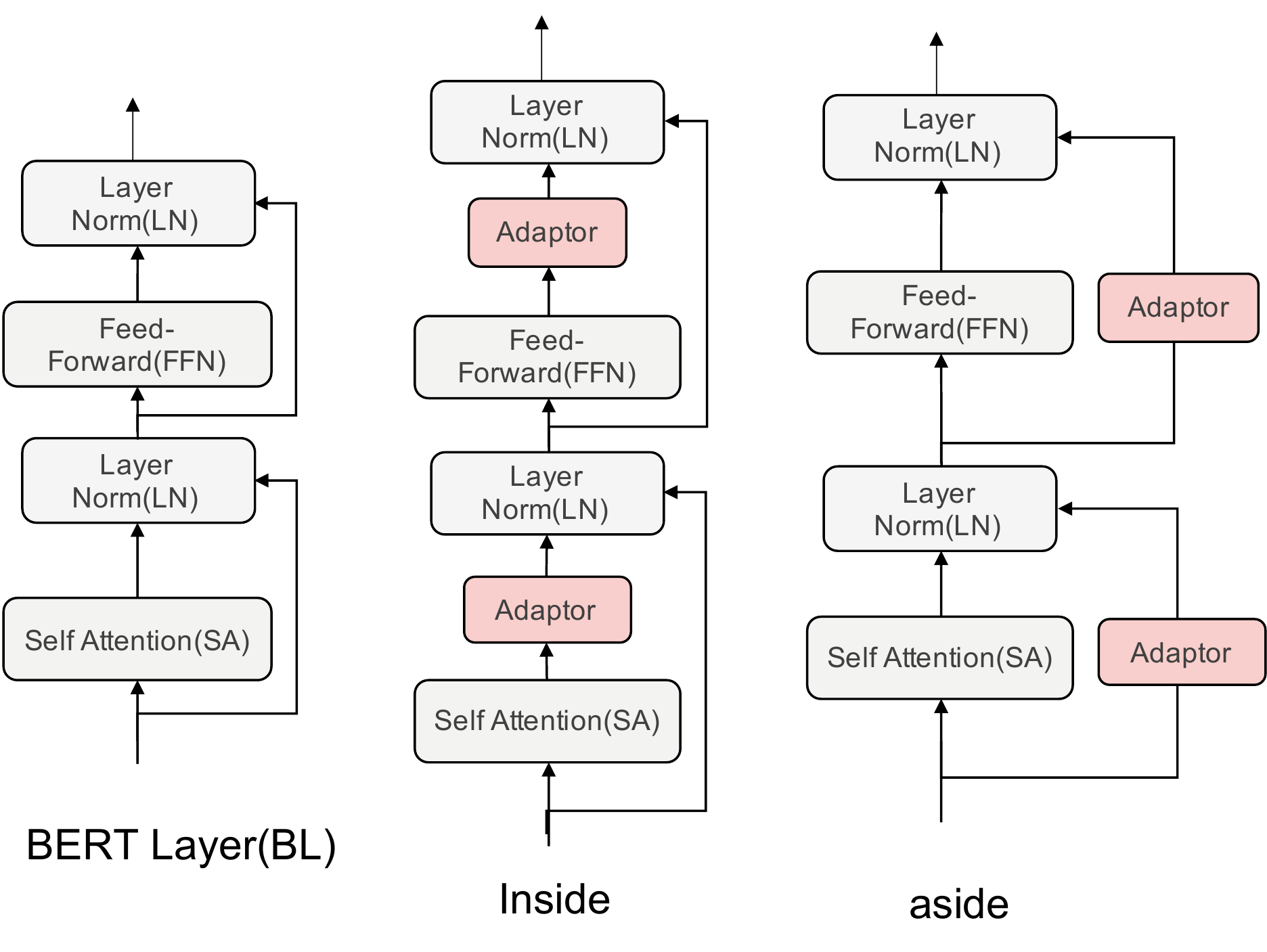}
 \caption{BERT Layer and two variants of adapted BERT Layer.}
 \label{fig:pos}
 \end{figure}

\subsection{ProgBERTQA}

Based on the dynamic-architecture paradigm, we propose a novel ProgBERTQA model for MRC, which enlarges its capacity when a new domain arrives.

\subsubsection{Overview}

Although modeling each domain separately could prevent catastrophic forgetting, it is impossible to support transfer learning which may lead to intractable model parameters. 
Also, sharing all parameters across domains is vulnerable to catastrophic forgetting as its fixed model capacity. 
Thus, ProgBERTQA shares common parameters among all domains and maintains domain-specific parameters to prevent forgetting for CDA. We denote the domain-specific parameters as adaptors, as it adapts original BERT outputs to different domains.

Specifically, ProgBERTQA is a combination of BERTQA and domain-specific parameters denoted as adaptors. 
BERT is used as the backbone structure for storing the common parameters, while each adaptor is explicitly applied to each incoming domain. Parameters in BERT are not updated during training, as BERT with pre-trained parameter has the ability to represent common sentences. An adaptor is responsible for adapting exiting BERT representation to the specific domain. New Adaptor for the subsequent domain is initialized by previously learned parameters of an adapter. In fact, storing and transferring knowledge for deep learning can be done in a straightforward manner through the learned weights. We thus use the learned adapter from the last domain to initialize the adapter for the current domain. In this way, we used the learned parameter as prior for the current domain adapter to forward transfer knowledge.

The model size of ProgBERTQA is 
$$N_{BERT}+T*N_{adapter},$$
where $N_{BERT}$ is the amount of BERT parameters, $N_{adapter}$ is the amount of adapter parameter, T is the number of domains. 
As the model size grows linearly with the number of domains, how to design an effective adapter which could achieve better performance with limited parameters is the core problem in ProgBERTQA. 
We now describe different ways of adapter insertion and adapter structure.

 \begin{figure}[t]
 \centering
 \includegraphics[scale=0.34]{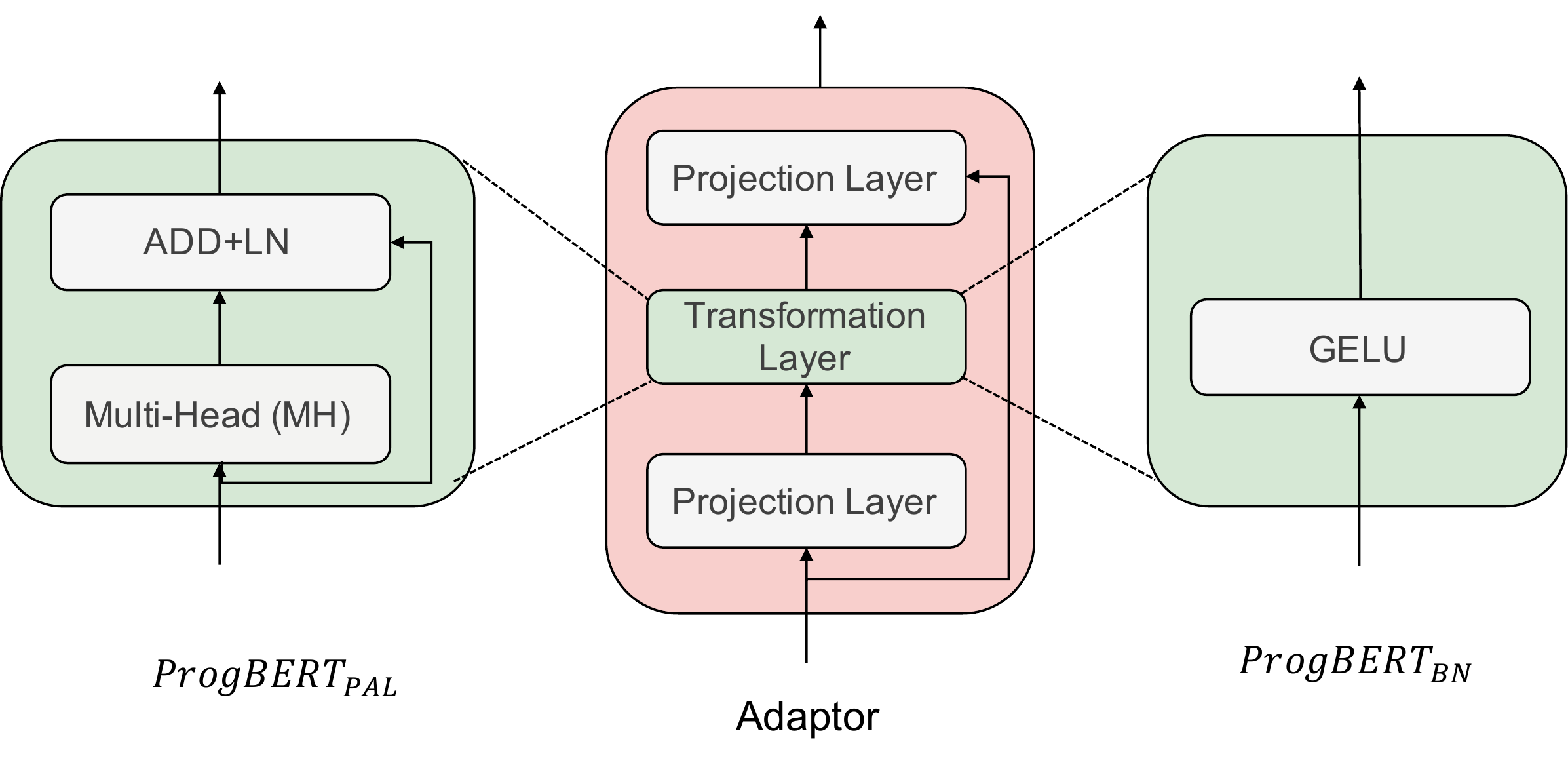}
 \caption{Two types of adapter structures.}
 \label{fig:fn}
 \end{figure}

\subsubsection{The Adapter Insertion}
In this section, we discuss where to insert the adapter into the BERTQA model.

Adding the adapter to each BERT layer is more parameter efficient than that to the topmost layer \cite{Stickland2019BERTAP}, i.e., achieving more effective capacity with the same number of parameters. 
Based on the previous introduction, each BERT layer contains a self-attention and a feed forward layer. We then add the adapter to these two components.
As depicted in Figure \ref{fig:pos}, we explore two position to insert adapter, namely, \emph{inside} and \emph{aside}. In  \emph{inside} fashion, as shown in Figure \ref{fig:pos} \emph{inside}, we put an adapter following the self attention component, another following the feed forward layer.
$$SA(h) = LN(Adapter(MH(h))+h),$$
$$BL(h) = LN(Adapter(FFN(SA(h))) +SA(h)),$$
where $Adapter(\cdot)$ denotes the adapter layer.
In \emph{aside} fashion, as shown in Figure \ref{fig:pos} \emph{aside} , we add two adapters parallel with multi-head attention and feed forward layer respectively.  
$$SA(h) = LN(MH(h)+Adapter(h)+h),$$
$$BL(h) = LN(FFN(SA(h))+ Adapter(SA(h)) +SA(h)).$$
\emph{Inside} fashion use adapter to reform the outputs of multi-head attention and feed forward layer.   
\emph{Aside} fashion use adapter  outputs as addend to  the outputs of multi-head attention and feed forward layer.

\subsubsection{The Adapter Structure}
In this section, we discuss how to design the adapter structure.

There are many choices on the adapter structure and we introduce two structures based on the attention and feed-forward layer respectively. 
The reason is that they have achieved a more effective model capacity with same number of parameters on multi-task learning \cite{Stickland2019BERTAP} and transfer learning \cite{Houlsby2019ParameterEfficientTL}.
ProgBERTQA aims to efficiently enlarge its capacity, thus we investigate these two structures in CDA for MRC. In the next section, we describe these two structures in detail.

As depicted in Figure~\ref{fig:fn}, both these two structures are surrounded by two projection layers. The first projection layer transforms the origin space of $d$ dimension to the space of $d_s$ dimension to save the budget of parameters. The output of the first projection is passed to a transformation layer. Finally, another projection layer transforms the vector to the space of $d$ dimension. These two structures are only different in the transformation layer. We list them as follows.
\begin{itemize}[leftmargin=*]
\item Inspire by \cite{Stickland2019BERTAP}, a multi-head attention layer is utilized as the transformation layer. The intuition behind this AL is that different domain needs different interaction between the token representations.
\item Inspire by \cite{Houlsby2019ParameterEfficientTL}, a non-linear activation function gelu \cite{Hendrycks2016GaussianEL} is used as transformation layer. The whole adapter is a bottleneck architecture. 
\end{itemize}
We denote these two adapter structures as Projected Attention Layer (PAL) and BottleNeck (BN) respectively.

We can see that the number of parameters in PAL is $3d_sd_s+2d_s d$ and the number of parameters in BN is $2d_sd$.  In order to fairly compare these two structure we keep the number of parameters in each structure the same. For example when $d_s$ in BN is 256, then the $d_s$ in PAL is set to 192.

By combining these two ways of adapter insertion (i.e., inside and aside) and two adapter structures (i.e., PAL and BN), we obtain four types of ProgBERTQA models denoted as 
ProgBERTQA$_{I-PAL}$, ProgBERTQA$_{I-BN}$, ProgBERTQA$_{A-PAL}$, ProgBERTQA$_{A-BN}$.

\subsection{Model Training and Testing}
In the training phase, we sequentially training the RegBERTQA and ProgBERTQA on each domain. For RegBERTQA, we update all the parameters. For ProgBERTQA, we update only the adapter and the MRC output layer.  We do not update the shared BERT parameter and keep them unchanged. The training loss of ProgBERTQA is the same as that of BERTQA~\cite{Devlin2019BERTPO}.

 \begin{table*}[t]
\centering
\renewcommand{\arraystretch}{1.5}
   \setlength\tabcolsep{10pt}
\begin{tabular}{cccccccccc}
\hline \hline
\textbf{} & what    & which   & where   & when    & how     & why     & other   & who     & overall  \\ \hline
BASE     &54.81&	57.95&	54.66&	65.29&	60.96&	46.23&	57.48&	81.4&	478.78 \\
RegBERTQA       &69.08&	64.98&	64.13&	71.33&	67.63&	59.93&	65.16&	77.66&	539.9\\
ProgBERTQA$_{A-PAL}$  &52.56&	61.88&	65.35&	74.85&	70.34&	57.28&	60.42&	80.89&	523.57 \\ 
ProgBERTQA$_{A-BN}$  & 63.03&	66.22&	69.91&	78.62&	72.26&	65.3&	66.37&	83.01&	564.72  \\ 
ProgBERTQA$_{I-PAL}$  &57.01&	53.39&	65.6&	75.12&	64.43&	53.46&	59.69&	79.51&	508.21 \\ 
ProgBERTQA$_{I-BN}$  &63.15&	70.47&	67.89&	79.23&	71.98&	69.47&	65.29&	83.28&	\textbf{570.76} \\ \hline \hline
\end{tabular}
\caption{Performance of models on CDA-Q.}
\label{tab:main_qtype}
\end{table*}

\begin{table*}[t]
\centering
\renewcommand{\arraystretch}{1.5}
   \setlength\tabcolsep{18pt}
\begin{tabular}{ccccccc}
\hline \hline
         & wiki    & news    & scripts & book    & tweet   & overall           \\ \hline
BASE   &67.9&	44.69&	67.8&	58.51&	85.12&	324.02 \\
RegBERTQA      &  79.46&	48.98&	62.8&	60.15&	84.76&  336.15              \\
ProgBERTQA$_{A-PAL}$   & 64.59&	49.34&	62&	56.6&	80.88&	313.41 \\
ProgBERTQA$_{A-BN}$   & 72.13&	56.06&	66.33&	60.98&	82.94&	338.44 \\
ProgBERTQA$_{I-PAL}$   & 67.14&	50.31&	63.09&	59.46&	83.12&	323.12 \\
ProgBERTQA$_{I-BN}$   & 73.66&	56.9&	67.86&	64.1&	85.4&	347.92\\ \hline \hline

\end{tabular}
\caption{Performance of models on CDA-C.}
\label{tab:main_ptype}
\end{table*}

In the testing phase, we test the model on all its seen domain. In ProgBERT model, domain label is needed, which indicates which adapter is activated to predict the results.

\section{Experiments}
In this section, we conduct experiments to verify the effectiveness of our proposed models.

\subsection{Experimental Settings}
We implement our models in PyTorch~\cite{NEURIPS2019_9015} based on Transformers library \footnote{https://github.com/huggingface/transformers}. We optimize the model using Adam \cite{Kingma2015AdamAM}  with warmup technique, where the learning rate is increased over the first 10\% of batches and then decayed linearly to zero. All runs are trained on Tesla V100 GPU with batch size 16. We train the model for each domain for 3 epochs and use the result from the last epoch.  The learning rate is tuned amongst $\{2e^{-5}, 5e^{-5}\}$. We use the base-uncased version of BERT and other related pre-trained BERT-related models can also be tried, such as RoBERTa \cite{Liu2019RoBERTaAR}, spanBERT \cite{Joshi2020SpanBERTIP}. The maximum sequence length of BERT  is set to 512. Datasets and codes are available at https://github.com/lixinsu/CDA-CIKM2020.

\subsection{Evaluation Metrics} \label{sec:eval}
Inspired by \cite{asghar2018progressive}, we report the answer accuracy of  each domain. We also show the overall performance of all domains by the sum of the accuracy in each domain.

Predicting answers accuracy is evaluated using exact match score (EM) and word-level F1-score (F1), as is common in extractive question answering tasks~\cite{Chen2017ReadingWT}. EM equals 1 only for the prediction that exactly matches the golden answer. F1 gives partial credit for the word overlap with the golden answer. We mainly use F1 as the evaluation criterion as it is more fine-granularity. For the implementation, we use the script from MRQA \footnote{https://github.com/mrqa/MRQA-Shared-Task-2019}.

\subsection{Models}

\subsubsection{Our Models}
We presents our proposed methods based on regularization and as follows:
\begin{itemize}[leftmargin=*]
\item \textbf{RegBERTQA}: Add a regularization term to prevent forgetting.
\item \textbf{ProgBERTQA$_{I-PAL}$}: Inserting projected attention layer inside each of the BERT layers. 
\item \textbf{ProgBERTQA$_{I-BN}$}: Inserting bottleneck layer inside each of the BERT layers.
\item \textbf{ProgBERTQA$_{A-PAL}$}: Inserting projected attention layer inside each of the BERT layers. 
\item \textbf{ProgBERTQA$_{A-BN}$}: Inserting bottleneck layer inside each of the BERT layers.
\end{itemize}

\subsubsection{Baselines}
We compare our methods with a naive adaptation of BERTQA. We denote the method which continually finetunes the BERTQA model as \textbf{BASE} method.

\subsection{Major Results}

We compare proposed ProgBERT models with existing continual learning baselines and the results are shown in Table~\ref{tab:main_qtype} and Table \ref{tab:main_ptype}. 
We have the following findings:
(1) From Table \ref{tab:main_qtype}, we can observe that regularization-based approach RegBERTQA achieves significantly better results than BASE methods. The gap in the overall score is about 44, and thus this demonstrates the effectiveness of the regularization-based method on BERTQA. Through compare performance of each domain performance, we observe that RegBERTQA achieves better performance on earlier domains and in the last domain it is worse than the BASE method. 
This indicates that the regularization-based method indeed hurt the modeling capacity of BERT and prevent the learning from new domains.
(2) Second, when comparing four variants of ProgBERT,  $ProgBERTQA_{I-*}$ outperform $ProgBERTQA_{A-*}$. This validates that inserting adaptor in the BERT layer is more effective than aside fashion. We also observe that  $ProgBERTQA_{*-BN}$ outperforms $ProgBERTQA_{*-PAL}$, this demonstrates two layer attention is better than projected attention layer.  $ProgBERTQA_{*-BN}$ achieves the best result among these four methods. In next 
section, we use $ProgBERTQA_{I-BN}$ for analysis and denoted as ProgBERTQA for brevity.
(3) When compare $ProgBERTQA_{I-BN}$ with RegBERTQA, we  can observe that $ProgBERTQA_{I-BN}$ outperforms RegBERTQA by a large margin. In Table \ref{tab:main_qtype}, except the first domain, $ProgBERTQA_{I-BN}$  beats  $ProgBERTQA_{I-BN}$ on all other domains.  This demonstrate the effectiveness of our proposed ProgBERT method.

\subsection{Breakdown Analysis}

Beyond above overall performance analysis, we also take some breakdown analysis for the CDA task.

\subsubsection{Assessing Catastrophic Forgetting}
 \begin{figure}[t]
 \centering
 \includegraphics[scale=0.45]{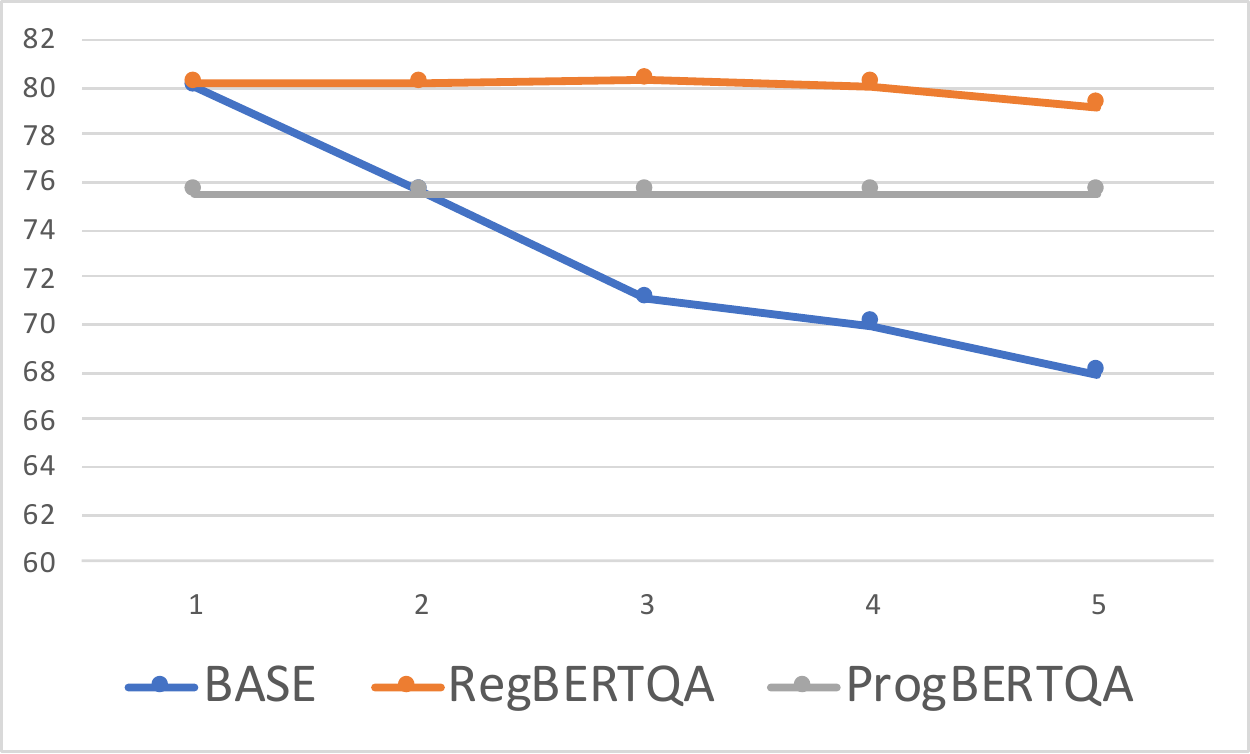}
 \caption{Catastrophic forgetting phenomenon of first domain.}
 \label{fig:cf1}
 \end{figure}

\begin{figure}[t]
 \centering
 \includegraphics[scale=0.5]{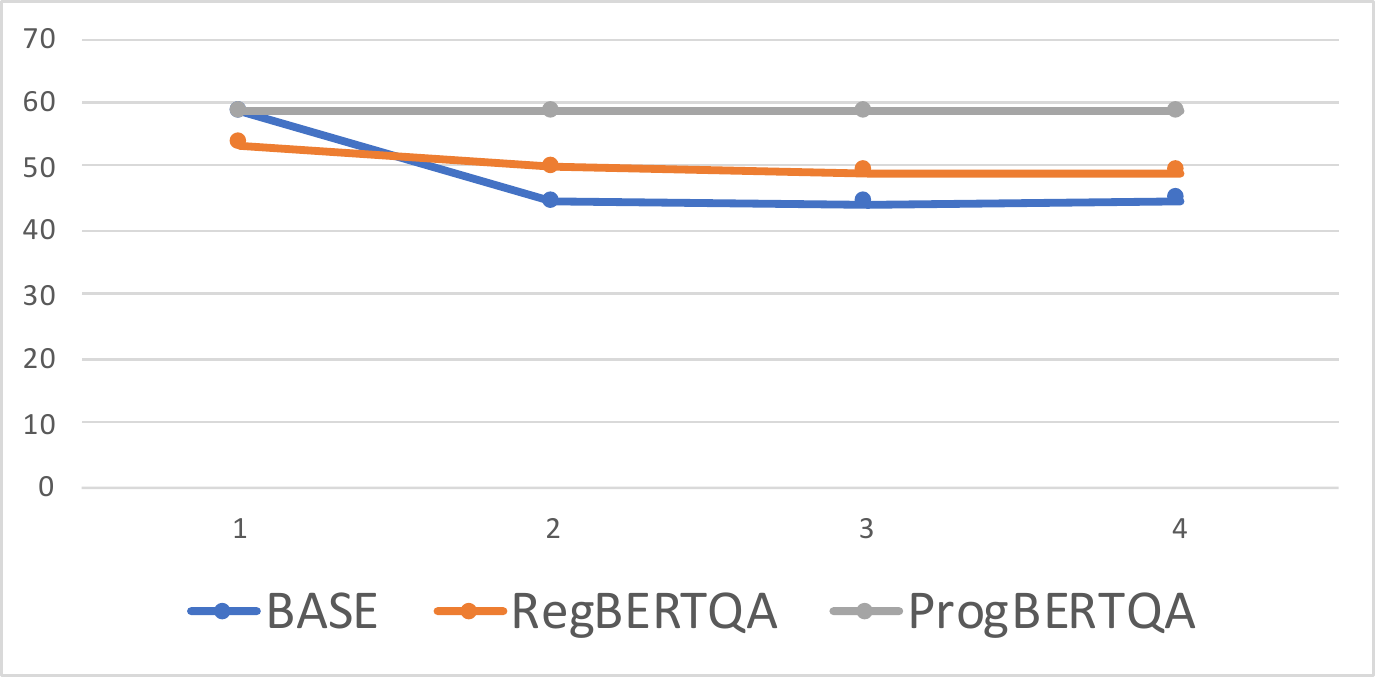}
 \caption{Catastrophic forgetting phenomenon of second domain.}
 \label{fig:cf2}
 \end{figure}

To assess catastrophic forgetting of proposed methods, we show how the performance of the first domain and second domain varies over the training process on the remaining domains \cite{Adel2020ContinualLW,Schwarz2018ProgressC}.  We conduct analysis experiment on CDA-C dataset.  We  display the results of first domain in Figure \ref{fig:cf1}. Note that we choose the $ProgBERTQA_{I-BN}$  and denote it as ProgBERTQA for brevity. We have following observation:
(1) We can see that ProgBERTQA can completely avoid catastrophic forgetting. The reason is that ProgBERTQA fixes the parameters of BERT and  domain-specific adapter.  However there is no free lunch, ProgBERTQA perform worse than RegBERTQA, when learning the first domain, as the fixed BERT limited the model capacity.
(2) From the Figure \ref{fig:cf1} and  \ref{fig:cf2}, we can observe that RegBERTQA still suffer from the catastrophic forgetting. The intuition is that RegBERTQA is with fixed model capacity which theoretically cannot capture all domains.
(3) Both ProgBERTQA and RegBERTQA outperform the BASE method, this demonstrate the effectiveness of our proposed two method for CDA for MRC.  

\subsubsection{Assessing Forward Transfer}

\begin{table*}[t]
\centering
\renewcommand{\arraystretch}{1.4}
   \setlength\tabcolsep{15pt}
\begin{tabular}{cccccccccc}  \hline \hline
dim & what    & which   & where   & when    & how     & why     & other   & who     & overall  \\ \hline
 32  & 54.20 & 55.66 & 65.44 & 74.83 & 69.32 & 64.07 & 57.98 & 80.18 & 521.73 \\
64  & 58.84 & 56.73 & 66.31 & 76.21 & 69.59 & 66.73 & 62.10 & 80.45  & 537.01 \\
128 & 58.92 & 65.45 & 68.54 & 77.51 & 69.68 & 64.54  & 65.49 & 82.80 & 552.96 \\
256 & 63.14 & 70.46 & 67.88 & 79.22 & 71.98 & 69.46  & 65.28 & 83.27 & \textbf{570.74} \\
384 & 64.74 & 68.98 & 70.74 & 79.31 & 73.19 & 62.30 & 66.11 & 83.40 & 568.79 \\
512 & 64.94 & 65.58 & 71.14 & 78.99 & 74.83 & 63.05 & 67.00 & 83.84 & 569.41  \\ \hline \hline
\end{tabular}
\caption{Performance of ProgBRETQA for different adapter size on CDA-Q.}
\label{tab:size_qtype}
\end{table*}

\begin{table}[t]
\centering
\renewcommand{\arraystretch}{1.5}
   \setlength\tabcolsep{3pt}
\begin{tabular}{lllllll}
\hline \hline
method                              & wiki & news & scripts & book & tweet & overall \\ \hline
ProgBERTQA & 75.52 & 58.63  & 70.51 & 66.62       & 86.31   & 357.59  \\
ProgBERTQA$_{init}$                              & 75.52 & 56.17  & 64.51 & 59.66       & 79.81   & 335.67  \\
INDIVIDUAL               & 80.01 & 58.14  & 66.27 & 58.27       & 83.01   & 345.7  \\ \hline \hline
\end{tabular}
\caption{Forward transfer analysis.}
\label{tab:transfer}
\end{table}

In this section, we explore whether ProgBERTQA model can transfer knowledge from the old domain to the new domain.
We denotes ProgBERTQA without initialization form previous domain as ProgBERTQA$_{init}$. We denotes individually fine-tune BERTQA on each domain as INDIVIDUAL. The results are shown in Table \ref{tab:transfer}. 
(1) We can see that ProgBERTQA outperform INDIVIDUAL in the last four domains. The reason may be INDIVIDUAL learns from each domain and some domain has small size of data. This demonstrate the ProgBERTQA can transfer knowledge from the old domain to the new domain.
(2) When we remove adapter initialization from last checkpoint, ProgBERTQA$_{init}$ is worse than ProgBERTQA. This demonstrate the necessity of prior initialization in ProgBERTQA, i.e., initializing new adaptors from last adaptors. 
%
%

\subsubsection{Assessing Order Robustness}
In this section, we compare the robustness of each method on the random order of domains. We permute the order of domain and test RegBERTQA and ProgBERTQA. 
The results are shown in Table \ref{tab:order}. order1 is a random order of domains in CDA-Q, order2 is an ascending order based on the number of training sample in each domain and order3 is a descending order. 

\begin{table}[t]
\renewcommand{\arraystretch}{1.3}
   \setlength\tabcolsep{12pt}
\centering
\begin{tabular}{cccc}
\hline \hline
  &  RegBERTQA & ProgBERTQA\\\hline
order1&    539.9  &      570.76             \\
order2&    401.35    &       542.97            \\
order3&     519.66   &        577.58           \\ \hline \hline
\end{tabular}
\caption{Order robustness analysis of RegBERTQA and  ProgBERTQA on CDA-Q.}
\label{tab:order}
\end{table}

From the results we have two observations:
(1) Through the performance variance between different orders, we can find that ProgBERTQA is more robust than RegBERTQA.
(2) The performance of RegBERTQA  degrades severely in order2, as order2  is an ascending order. This is because RegBERTQA is poisoned by domain which has insufficient data or corrupted labels, so RegBERTQA converges to a bad point in parameter space. RegBERTQA cannot escape from that bad  
converge point, and thus this may cause the order-sensitivity of the RegBERTQA model.
(3) ProgBERTQA is more robust compared to RegBERTQA, since it only initializes the adapter with the previous parameters.
In conclusion, when there are low-quality data in specific domains, ProgBERTQA is more preferable then RegBERTQA.

\subsubsection{Analysis of the Effect of Adapter Size}
Adaptor size is a hyper-parameter in our proposed ProgBERTQA model. 
Smaller adapter consumes fewer parameters while the performance may decrease. 
In this section, we test the performance over different adaptor size. We tune the adapter size, i.e., the size of the dimension of projection space $d_s$. We vary the adaptor size in $\{32,64,128,256,384,512\}$ and conduct experiments on both CDA-Q and CDA-C. The results are shown in Table \ref{tab:size_qtype} and \ref{tab:size_ctype}. 
From the results we can see that the 256 is the best adaptor size for CDA-Q, and 512 for CDA-C.
From Table \ref{tab:size_qtype}, we can see that when the adapter size exceeds some threshold, the performance gets worse as the adapter gets bigger.  A possible reason is the overfitting of adapter. 
There is a trade-off between the adapter size and the performance. We can control the adapter size to balance the limitation on the number of parameters and the performance requirement.

\begin{table}[t]
\renewcommand{\arraystretch}{1.3}
   \setlength\tabcolsep{5pt}
\centering
\begin{tabular}{ccccccc}
\hline \hline
dim  & wiki    & news    & scripts & book    & tweet   & overall  \\ \hline 
 32  & 65.79&	51.6 &	63.41&	60.33&	82.73&	323.86 \\
64  & 69.32&	53.93&	65.25&	61.19&	83.47&	333.16 \\
128 & 72.41&	55.87&	66.77&	63.41&	84.11&	342.57\\
256 &73.66&	56.9&	67.86&	64.1&	85.4&	347.92  \\
384 &75.36&	58.03&	69.99&	65.49&	84.8&	353.67 \\
512 & 75.52&	58.63&	70.51&	66.62&	86.31&	\textbf{357.59}  \\ \hline \hline
\end{tabular}
\caption{Performance of ProgBRETQA for different adapter size on CDA-C.}
\label{tab:size_ctype}
\end{table}

\section{Conclusion}
We introduce the \textit{Continual Domain Adaptation} task for MRC. So far as we know, this is the first study on the continual learning perspective of MRC. We build two datasets CDA-Q and CDA-C for the CDA task, by re-organizing existing MRC collections into different domains with respect to the  question type and passage type. 
We conduct preliminary experiments showing the existence of catastrophic forgetting (CF) phenomenon of existing MRC models under the CDA setting. 
Further, we propose regularization-based RegBERTQA and dynamic-architecture ProgBERTQA to tackle the CDA for MRC. 
We conduct extensive experiments to analysis the effectiveness of both methods and validate that the proposed dynamic-architecture based model achieves the best performance.

\begin{acks}
This work was supported by Beijing Academy of Artificial Intelligence (BAAI) under Grants No. BAAI2019ZD0306 and BAAI2020ZJ0303, and funded by the National Natural Science Foundation of China (NSFC) under Grants No. 61722211, 61773362, 61872338, and 61902381, the Youth Innovation Promotion Association CAS under Grants No. 20144310, and 2016102, the National Key RD Program of China under Grants No. 2016QY02D0405, the Lenovo-CAS Joint Lab Youth Scientist Project, the K.C.Wong Education Foundation, and the Foundation and Frontier Research Key Program of Chongqing Science and Technology Commission (No. cstc2017jcyjBX0059).
\end{acks}

\bibliographystyle{ACM-Reference-Format}
\bibliography{sample-base}


\end{document}